\documentclass[trackchanges,twocolumn]{aastex701} 

\usepackage{color}

\begin{document}

\title{Mass-Ratio Reversal as an Alternative to Hierarchical Mergers for GW241011}

\author[orcid=0000-0002-6442-7850]{Rui-Chong Hu}
\affiliation{Nevada Center for Astrophysics, University of Nevada, Las Vegas, NV 89154, USA; \url{ruichong.hu@unlv.edu}}
\affiliation{Department of Physics and Astronomy, University of Nevada, Las Vegas, NV 89154, USA}
\email{ruichong.hu@unlv.edu}  

\author[orcid=0000-0002-2956-8367]{Ying Qin} 
\affiliation{Department of Physics, Anhui Normal University, Wuhu, Anhui, 241002, China; \url{yingqin@ahnu.edu.cn}}
\email{yingqin@ahnu.edu.cn}

\author[orcid=0000-0002-9725-2524]{Bing Zhang}
\affiliation{The Hong Kong Institute for Astronomy and Astrophysics, The University of Hong Kong, Pokfulam Road, Hong Kong, China; \url{bzhang1@hku.hk}}
\affiliation{Department of Physics, Department of Physics, The University of Hong Kong, Pokfulam Road, Hong Kong, China}
\affiliation{Nevada Center for Astrophysics, University of Nevada, Las Vegas, NV 89154, USA; \url{ruichong.hu@unlv.edu}}
\affiliation{Department of Physics and Astronomy, University of Nevada, Las Vegas, NV 89154, USA}
\email{bzhang1@hku.hk}

\begin{abstract}

Recent gravitational-wave (GW) observations have revealed binary black hole (BBH) mergers with both extreme mass ratios and large effective spin parameters ($\chi_{\rm{eff}}$). GW241011 is a notable example that shows these properties. Although hierarchical mergers (second-generation + first-generation BHs) can naturally produce high spins and become more prevalent than first-generation + first-generation mergers at mass ratios $q \lesssim 0.6$, they are negligible at the extreme mass ratio $\sim 0.3$ and are further limited by gravitational recoil kicks that can eject the second-generation BH from the host environment. Moreover, recent studies have argued against a dynamical origin for GW241011. Here, we investigate the formation of GW241011-like systems through the mass-ratio reversal (MRR) channel in isolated binary evolution. By quantifying the probability of producing such systems across a range of binary-evolution models, we identify the key dependencies on stellar-evolution and binary-interaction physics. Our results demonstrate the conditions under which the MRR channel can provide a viable alternative to hierarchical mergers for GW241011-like events, although its predicted merger rate density remains relatively low.

\end{abstract}

\keywords{\uat{Binary stars}{154} --- \uat{Gravitational waves}{678} --- \uat{Stellar evolution}{1599}}

\section{Introduction}\label{intro}

During the second part of the fourth observing run (O4b) of the LIGO-Virgo-KAGRA (LVK) Collaboration network, two gravitational-wave (GW) signals from binary black hole (BBH) coalescences were first reported: GW241011$_{-}$233834 and GW241110$_{-}$124123 \citep{GW241011}. Among these, GW241011$_{-}$233834 (hereafter abbreviated GW241011) is particularly remarkable for exhibiting the largest lower bound to date on the dimensionless spin magnitude of the more massive component (primary), with $\chi_1 > 0.69$ at 95\% credibility. The spin of the primary is inferred to be misaligned with respect to the Newtonian orbital angular momentum by $\theta = {31^{+11}_{-14}}\,^{\circ}$, while the secondary spin remains unconstrained. Parameter estimation further yields an effective inspiral spin of $\chi_{\rm eff} = 0.49^{+0.06}_{-0.07}$, together with a source-frame primary mass of $M_1 = 19.6^{+3.6}_{-2.5}$ $M_{\odot}$ and a significantly asymmetric mass ratio of $q = 0.30^{+0.09}_{-0.08}$, where $q = M_2/M_1$.

Hierarchical mergers in dense stellar environments provide a natural pathway for the formation of second-generation BHs, which can retain large spin magnitudes, populate the pair-instability mass gap, and participate in subsequent mergers that produce BBHs with highly asymmetric mass ratios, non-negligible spin-orbit misalignments, and even negative effective spins \citep[e.g.][]{Antonini2019,Rodriguez2019,Zhang2023}.
In such a scenario, repeated mergers in globular clusters, nuclear star clusters, or young massive star clusters \citep{Portegies2010,Neumayer2020} can therefore naturally give rise to BBHs with properties broadly consistent with those inferred for GW241011, including large spin magnitudes and unequal mass ratios. Moreover, clusters formed in relatively high-metallicity environments (e.g., $Z \gtrsim 0.1\, Z_\odot$, where $Z_\odot$ denotes solar metallicity) are capable of producing BBHs with primary masses comparable to that inferred for GW241011 \citep{Ye2026}.
However, this scenario faces several challenges. A recent study by  \cite{Vijaykumar2026}, based on the Cluster Monte Carlo \cite[CMC,][]{Kremer2020} simulations of dense star clusters, shows that 1 generation (G) + 2G mergers contribute more BBH mergers than 1G + 1G mergers at mass ratio $q \lesssim 0.6$, whereas their contribution becomes negligible at $q \sim 0.3$. Within the globular-cluster models with recoil calculations considered by \citet{Llobera2026}, their population analysis based on the adopted astrophysical models suggests that GW241011 would likely be ejected from typical globular clusters, even though retention in nuclear star clusters remains possible. In particular, gravitational-wave recoil velocities of merger remnants are typically of order a few hundred $\rm km\, s^{-1}$, with an extended high-velocity tail. Such large kicks can efficiently eject second-generation BHs from globular clusters, while retention in nuclear star clusters remains possible but is not guaranteed. Furthermore, their population analysis yields Bayes factors that strongly favor a field origin over a dynamical origin. In addition to hierarchical BH mergers, stellar mergers in young dense clusters have recently been proposed as another dynamical pathway for producing rapidly spinning BHs \citep{Satish2026}. However, it remains unclear whether this channel can simultaneously reproduce the large primary spin, spin-orbit misalignment, and extreme mass ratio observed in GW241011.

The high spin inferred for the primary BH in GW241011 is challenging to reconcile with the standard isolated binary evolution channel involving common-envelope (CE) evolution \citep[e.g.,][]{Tutukov1973,Phinney1991,Ivanova2013,Belczynski2016}. In this framework, the initially more massive star evolves off the main sequence first and eventually collapses to form the first-born BH (the primary). If angular momentum transport within massive stars is efficient—potentially mediated by magnetic torques \citep{Spruit1999,Spruit2002}—the stellar core is expected to be strongly coupled to the envelope and to lose most of its angular momentum prior to core collapse \citep{Qin2019}. Consequently, the nascent primary BH is predicted to be born with a low dimensionless spin parameter \citep{Qin2018,Fuller2019,Belczynski2020}. This expectation is at odds with the large primary spin inferred for GW241011 and more generally with the predictions of the standard CE channel \citep{Qin2022}. 

Motivated by the high effective spins inferred for several BBH mergers \citep{Abbott2021,Abbott2024}, \citet{Olejak2021} proposed two isolated binary evolution scenarios capable of producing rapidly spinning primary BHs. One scenario involves an equal-mass binary undergoing stable mass transfer followed by a CE phase, ultimately forming a close binary composed of two helium (He) stars. Strong tidal interactions efficiently spin up both He stars prior to core collapse, leading to the formation of two rapidly rotating BHs \citep[see details in][]{Qin2023}. In an alternative scenario, the initially more massive star in an unequal-mass binary system expands after leaving the main sequence and initiates mass transfer onto its companion during the first Roche-lobe overflow. In this phase, the accretor gains a substantial amount of mass and eventually becomes more massive than the donor, resulting in a reversal of the mass ratio. Mass-ratio reversal (MRR) has been invoked to explain a variety of interacting binary systems, including Algol binaries and compact binaries containing young neutron stars \citep[e.g.,][]{Portegies1996}, neutron star--black hole binaries \citep[e.g.,][]{Sipior2004}, and neutron star--white dwarf binaries \citep[e.g.,][]{Tauris2000,Toonen2018}. Observational support for MRR is provided by the presence of young neutron stars in neutron star--white dwarf binaries \citep[e.g.,][]{Tauris2000,Kaspi2000,Ng2018,Venkatraman2020}. In this scenario, mass accretion rejuvenates the initially less massive star \citep[e.g.,][]{Renzo2023,Qi2025,Xu2025}, allowing it to evolve as the more massive component and, if sufficiently massive, collapse to form the second-born BH. Owing to tidal spin-up in a close orbit prior to core collapse, this BH can be born with a high spin \citep{Detmers2008,Qin2018,Zaldarriaga2018,Bavera2020,Hu2022,Ma2023,Wang2026}. 
This scenario has been found across a range of population-synthesis simulations \cite[e.g.,][]{Zevin2022,Olejak2024,Chen2026,Smith2026}. In particular, \citet{Broekgaarden2022} conducted an extensive suite of population-synthesis simulations to investigate the properties of merging BBHs formed through MRR during the first RLOF phase. Their results suggest that MRR, resulting in the more massive BH forming second, is likely a common outcome in the evolutionary history of BBH mergers observed by current gravitational-wave detectors. Notably, \citet{Xu2025} recently pointed out that GW241011's high-spin primary is consistent with an MRR origin, although only within the stable mass transfer channel.

In this work, we perform binary population-synthesis calculations to investigate the formation of GW241011 through the MRR channel. The remainder of this paper is organized as follows. In Sect.~\ref{method}, we describe the population-synthesis method and the underlying physical assumptions adopted in our calculations. In Sect.~\ref{result}, we present the results of our main findings. Finally, in Sect.~\ref{con}, we summarize the conclusions with some discussion. 

\setcounter{footnote}{0}

\section{Methods}\label{method}
We perform binary population-synthesis calculations using the open-source code \texttt{COMPAS}\footnote{https://github.com/TeamCOMPAS/COMPAS} \citep[version v03.27.03;][]{Stevenson2017,Vigna2018,Neijssel2019,Riley2022,Disberg2025}, which employs parameterized prescriptions for single \citep{Hurley2000} and binary stellar evolution \citep{Hurley2002} to efficiently evolve large populations of binary systems. 
We adopt population synthesis models summarized in Table 1 of \cite{Hu2026}, which is based on the evolution framework of \cite{Broekgaarden2021}. 
We consider metallicities in the range $Z\in[0.0001, 0.03]$, adopting $Z_{\odot} = 0.0142$ \citep{Asplund2009}. For each of 10 metallicity bins spanning this range, we evolve $10^6$ binaries.

The stellar winds of massive stars, such as He stars, remain an open question in stellar evolution \citep{Vink2022}. 
In our fiducial model, we use metallicity-dependent stellar wind prescriptions, following the settings in \cite{vanSon2025} and \cite{Merritt2025}, based on the latest research development. 

During the stable Roche-lobe overflow, the fraction of mass lost by the donor that is accreted by the companion is parameterized by the mass-transfer efficiency, $\beta$, such that $\dot{M}_{\text{acc}} = -\beta \dot{M}_{\text{don}}$, where $0 \leq \beta \leq 1$. In the fiducial model, the efficiency is not fixed but is determined by the accretor's ability to accept mass on its thermal timescale \cite[see more details in][]{TeamCOMPAS2022}. To assess the impact of mass-transfer efficiency, we additionally consider models with fixed values of $\beta = 0.25,\,0.5$, and 0.75. We use the $\zeta-$prescription implemented in COMPAS to determine the stability of Roche-lobe overflow \citep{Vigna2018,TeamCOMPAS2022}. The stability is determined using approximations to the mass–radius relations \citep{Soberman1997}. Specifically, the donor's response to mass loss is described by analytic prescriptions for the mass--radius exponent, which depend on the donor's evolutionary state, and is compared with the response of the Roche-lobe radius to mass transfer. If the transferred mass exceeds the amount that can be accreted by the companion, the excess material is assumed to be ejected from the vicinity of the accretor via ``isotropic re-emission'', carrying away the accretor's specific orbital angular momentum \citep[e.g.][]{Bhattacharya1991,Tauris2006}. 

For CE evolution, we adopt the $\alpha-\lambda$ formalism \citep{Webbink1984,DeKool1990}. In this framework, the CE efficiency parameter ($\alpha$) quantifies the efficiency with which orbital energy is used to eject the envelope, while the parameter ($\lambda$) characterizes the binding energy of the envelope. In the fiducial model, $\lambda$ is calculated using the ``Nanjing lambda'' of \cite{Xu2010a,Xu2010b}, which provide fitting formulae based on detailed stellar evolution models. We adopt $\alpha_{\mathrm{CE}}=1$ as our fiducial value and additionally explore models with $\alpha_{\mathrm{CE}} = 0.5,\,2,\,5$ to assess the impact of CE efficiency on the formation of merging BBHs.

We adopt the ``delayed'' SN prescription \citep{Fryer2012} to determine compact object masses during CCSNe, which allows the production of BHs within the lower mass gap. If the ejecta are asymmetric, the remnant may receive a natal kick \citep[e.g.][]{Janka1994,Wongwathanarat2013}. We adopt the recent natal-kick prescription of \cite{Disberg2025} as our fiducial model, in which kick velocities follow a lognormal distribution with $\mu = 5.6$ and $\sigma = 0.68$. For comparison, we also consider a Maxwellian distribution with dispersions of $\sigma_{\mathrm{CCSN}}=100\,{\rm km}\,{\rm s}^{-1}$. We assume that the mass accretion onto the BH is limited by the Eddington rate.

From the simulated BBH population, we select systems in which the initially less massive star ultimately produces the more massive BH. We refer to such systems as MRR events and define the later-formed, more massive BH as the primary component, $M_1$.

We model the spins of the resultant BHs following the prescriptions of \cite{Bavera2020,Bavera2021}. The first-born BH is assumed to be non-spinning, consistent with efficient angular-momentum transport in its progenitor star \citep{Qin2018,Fuller2019}. The second-born BH, however, may be born with significant spin if its progenitor He star is tidally spun up in a sufficiently close binary prior to core collapse \citep{Qin2018,Zaldarriaga2018,Bavera2020,Hu2022,Ma2023}. This tidal spin-up can efficiently transfer orbital angular momentum to the He star and produce a rapidly rotating BH at birth.

We calculated the intrinsic merger-rate density by convolving the COMPAS formation efficiencies and delay-time distributions with a metallicity-specific cosmic star-formation history, following Equation (2) of \cite{Broekgaarden2021}. We adopted the star-formation-rate density from \cite{Madau2014}, the galaxy stellar-mass function from \cite{Panter2004}, and the mass–metallicity relation prescrived by \cite{Langer2006} with an offset from \cite{Broekgaarden2021}. We define GW241011-like systems as binaries whose parameters fall within the joint 90\% highest-posterior-density region of the kernel density estimation distribution in the $(M_1, q, \chi_{\rm eff})$ parameter space.

\section{Results}\label{result}

\begin{figure}
\centering
\includegraphics[width=0.95\linewidth, trim = 0 0 0 0, clip]{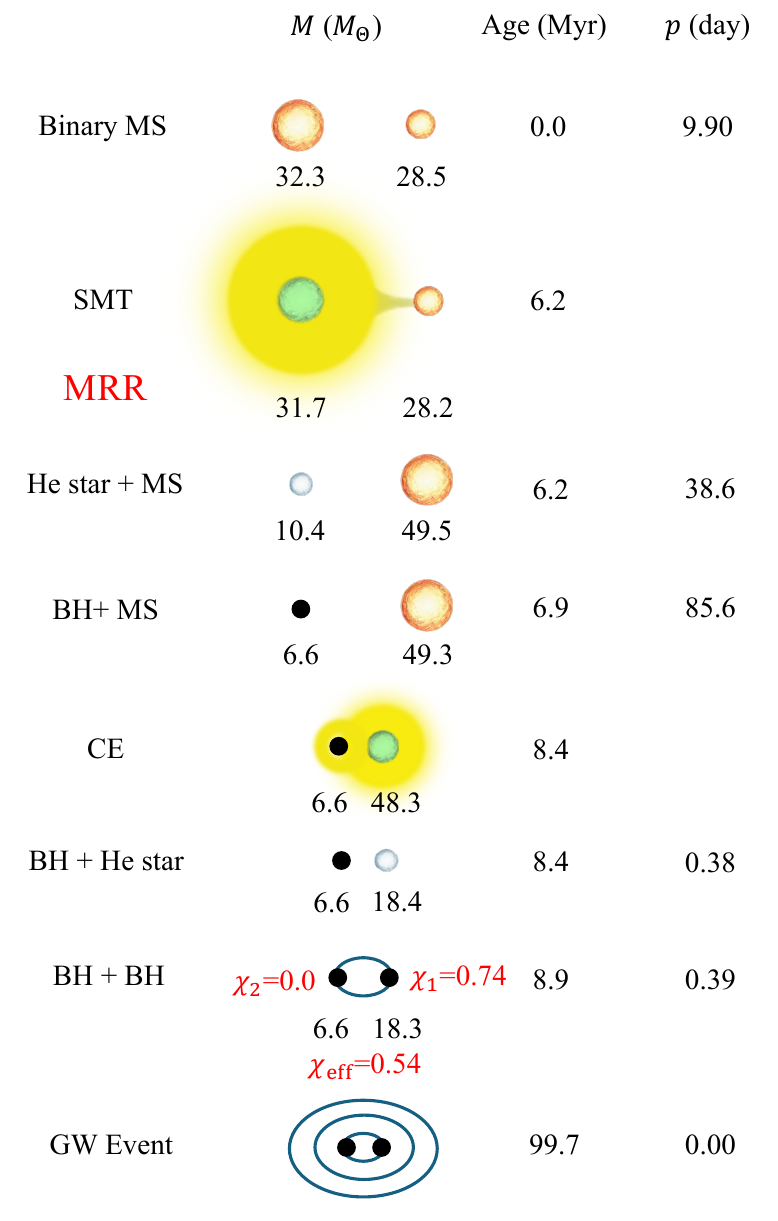}
\caption{Representative evolutionary pathway of the MRR channel leading to the formation of GW241011-like event with $\chi_1 = 0.74$. The system undergoes stable Roche-lobe overflow initiated by the initially more massive star, resulting in MRR. The initially less massive star later initiates a CE phase, whose ejection dramatically reduces the orbital separation. Following the CE phase, strong tidal interactions spin up the He-star progenitor of the second-born BH. The subsequent core collapse gives rise to a rapidly rotating BH, and the resulting BBH merges via gravitational-wave emission after a delay time of 99.7 Myr.
The underlying population adopts the fiducial mass-transfer prescription, lognormally distributed natal kicks with $(\mu=5.6)$ and $(\sigma=0.68)$, and $\alpha_{\rm CE}=5$.}
\label{fig:illustration}
\end{figure}

Fig.~~\ref{fig:illustration} illustrates a representative evolutionary pathway leading to the formation of GW241011 through the MRR channel. The binary initially consists of two massive stars in a relatively close orbit. As the primary star, initially the more massive component, evolves off the main sequence, it expands and eventually fills its Roche lobe, initiating mass transfer onto the secondary (initially less massive). 
During this Roche-lobe-overflow phase, mass transfer follows our fiducial prescription. The secondary retains enough of the transferred material to become more massive than the donor, thereby reversing the mass ratio.

After being stripped of its hydrogen-rich envelope, the initially more massive star undergoes core collapse and forms the first-born BH. The accretor, now the more massive component, subsequently evolves off the main sequence and initiates a CE phase. The successful ejection of the CE significantly shrinks the binary orbit, producing a tight BH--He star system. Following the CE phase, the orbital separation becomes sufficiently small for strong tidal interactions to operate efficiently. These tidal torques spin up the He star and can drive it toward synchronous rotation with the orbit, allowing it to acquire a substantial amount of angular momentum. When the He star eventually undergoes core collapse, it forms a rapidly rotating second-born BH. Owing to the earlier MRR, the second-born BH is more massive than the first-born BH and is therefore identified as the primary component of the BBH. In the example shown, the resulting BBH merger consists of a $18.3\,M_\odot$ primary BH, formed second, and a $6.6\,M_\odot$ secondary BH. The efficient tidal spin-up of the He star produces a rapidly rotating primary BH, yielding an effective inspiral spin parameter of $\chi_{\rm eff} = 0.54$, consistent with the high-spin nature of GW241011. This evolutionary channel naturally reproduces the two defining characteristics of GW241011: a rapidly rotating primary BH and a MRR that results in the more massive BH forming after its less massive companion.

\begin{figure*}
\centering
\includegraphics[width=0.98\linewidth, trim = 0 0 0 0, clip]{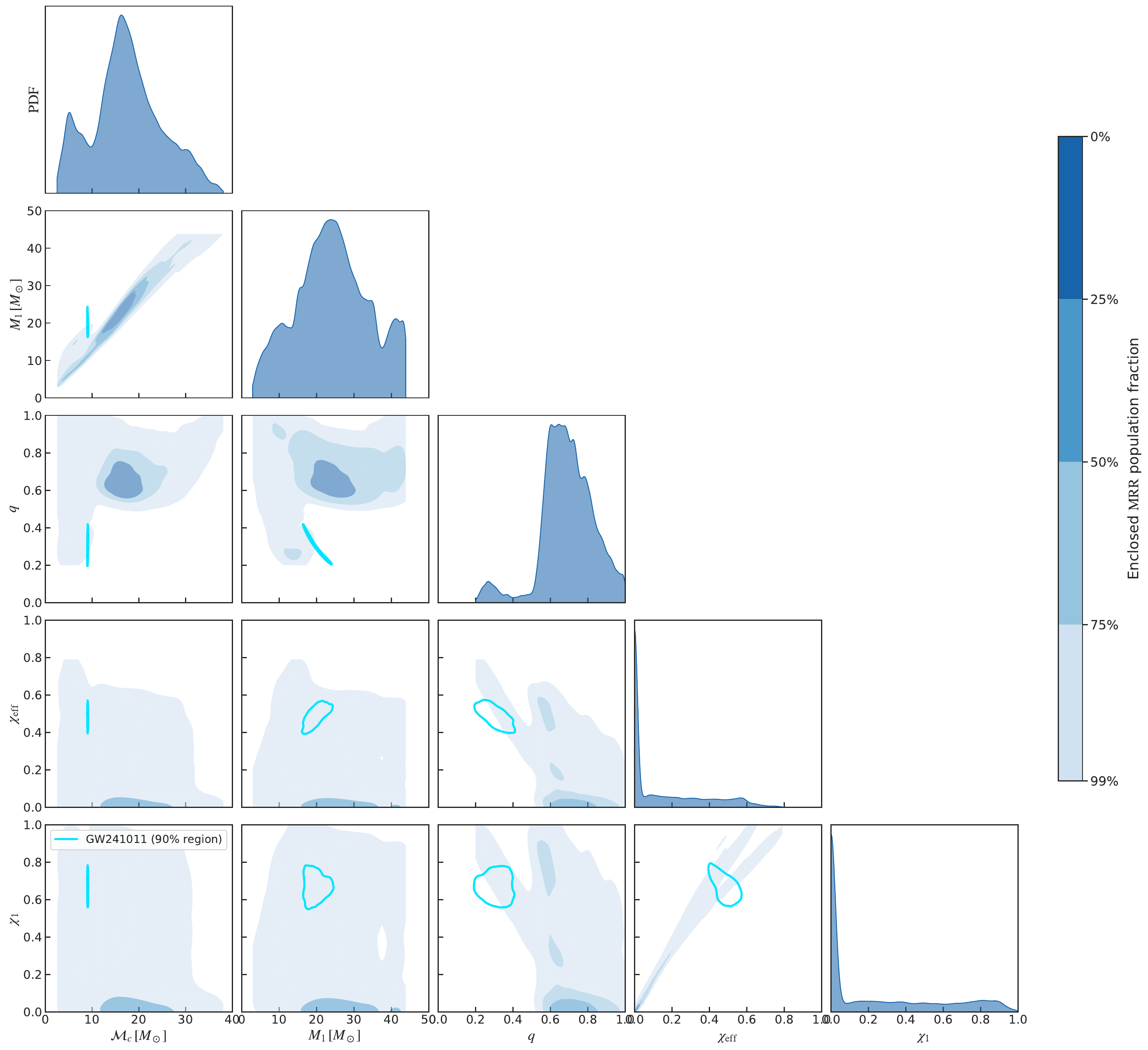}
\caption{Corner plot showing the 1D and 2D distributions of BBH properties produced through MRR channel of the same binary population synthesis model in Fig. \ref{fig:illustration}. The simulated BBH population is shown by light-to-dark blue filled contours, while the parameter space of GW241011 is shown in cyan (90\% credible regions).}
\label{fig:parameter_space}
\end{figure*}

Fig.~\ref{fig:parameter_space} presents the simulated population of MRR BBHs from our model adopting the fiducial setup except for $\alpha_{\rm{CE}}=5$ and the 90\% credible contours of the LVK inferred posterior samples for GW241011 \citep{GW241011}. We find that only a small fraction of MRR systems ($\sim 0.2\%$) overlap with the inferred posterior region of GW241011, and this overlap is highly sensitive to the adopted binary-evolution parameters. 
By adopting Madau–Panter–Langer cosmic-evolution prescription, we estimate a local ($z=0.05$) event rate density for GW241011-like systems of $2.5\times10^{-5}\, \rm{Gpc}^{-3}\, \rm{yr}^{-1}$ in the $\alpha_{\rm{CE}}=5$ model.
Notably, systems consistent with GW241011 are produced exclusively through the CE subchannel and only for a relatively high CE efficiency, $\alpha_{\rm CE} = 5.0$. For lower values of $\alpha_{\rm CE}$, the majority of progenitor binaries merge during the CE phase and fail to form merging BBHs, indicating that the survival of such systems requires highly efficient envelope ejection. Therefore, if GW241011 originated through the MRR channel in isolated binary evolution, our models preferentially reproduce its observed properties with a high effective CE efficiency (e.g., $\alpha_{\rm CE} = 5.0$). Large effective CE efficiencies have also been inferred from evolutionary reconstructions of massive BH X-ray binaries \citep{Li2026}. In the context of the adopted CE prescription, this may indicate that additional energy sources, such as ionization and thermal energy \citep[e.g.,][]{Fragos2019,Ivanova2013}, contribute to successful envelope ejection.

\begin{figure}
\centering
\includegraphics[width=0.95\linewidth, trim = 0 0 0 0, clip]{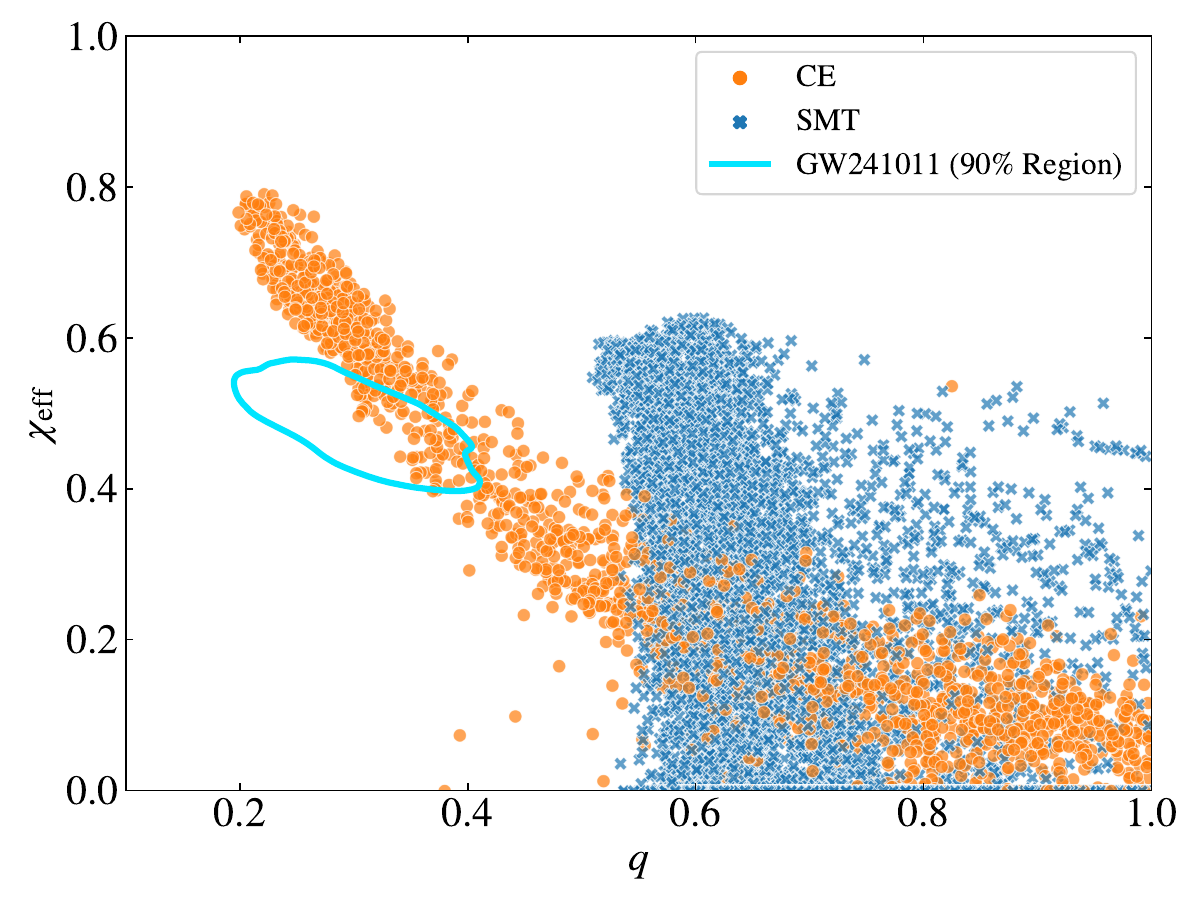}
\caption{Distribution of the effective spin parameter ($\chi_{\rm{eff}}$) as a function of the mass ratio ($q$) for MRR BBHs from the same model in Fig. \ref{fig:illustration}. The population is partitioned by binary evolutionary channel: SMT (blue) and CE (orange). The blue contour denotes the 90\% credible region inferred for GW241011.}
\label{fig:subchannel}
\end{figure}

To better understand the origin of these systems, we separate the MRR population into two subchannels: those involving a CE phase and those undergoing stable mass transfer (SMT), as shown in Fig.~\ref{fig:subchannel}. The CE channel produces BBHs spanning a wide range of mass ratios, including highly asymmetric systems ($q \sim 0.2$). It exhibits a pronounced anti-correlation between $q$ and $\chi_{\rm eff}$, with the most asymmetric binaries preferentially reaching the highest effective spins. In contrast, the SMT channel primarily forms binaries with more comparable component masses ($q > 0.5$). Moreover, because orbital contraction is less efficient in the SMT channel than in the CE channel, tidal spin-up of the progenitor stars is weaker, resulting in systematically lower effective inspiral spins. We note that the 90\% credible region of GW241011 falls exclusively within the region populated by CE systems and shows negligible overlap with the SMT population. We therefore infer that GW241011-like systems are formed predominantly through the CE subchannel, underscoring the importance of CE evolution in producing highly asymmetric BBHs.

\begin{figure}
\centering
\includegraphics[width=0.96\linewidth, trim = 0 0 0 0, clip]{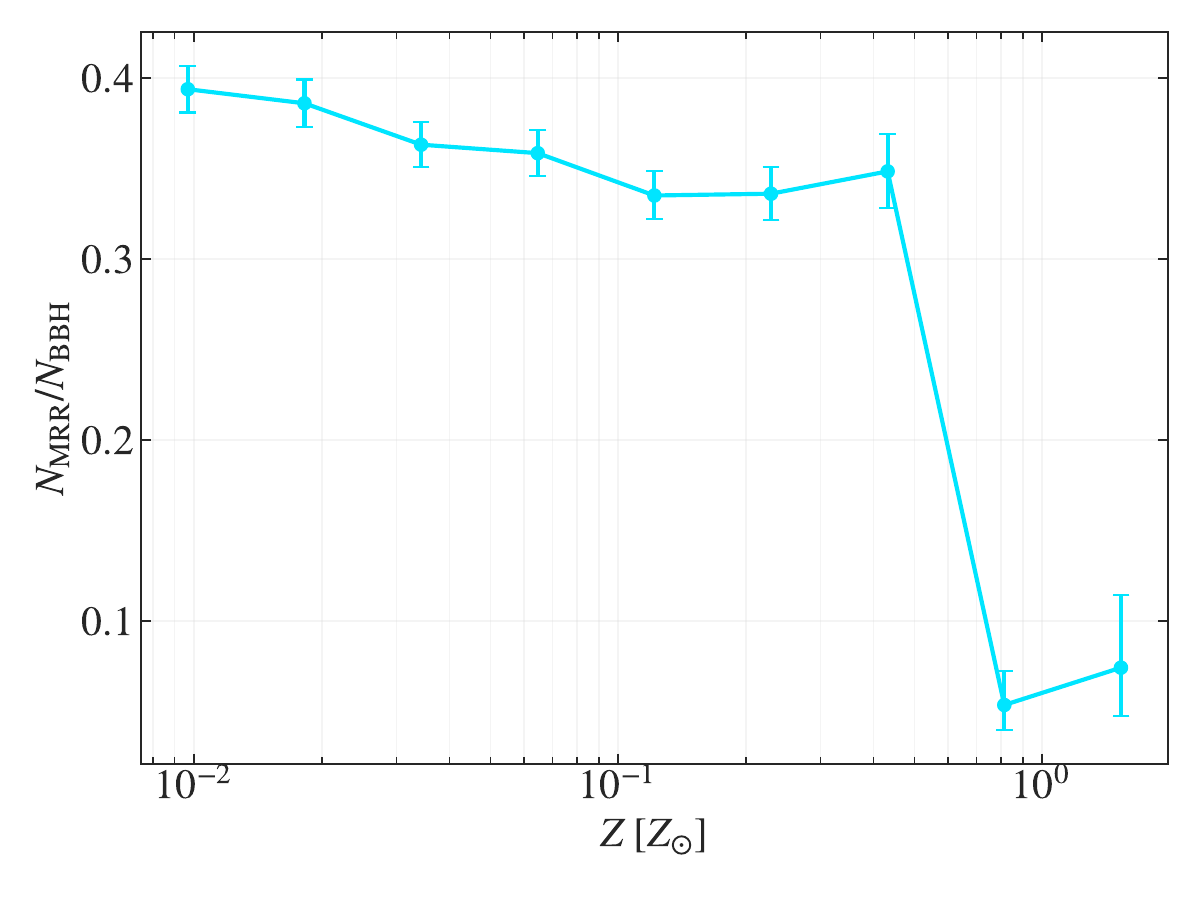}
\caption{Metallicity dependence of the MRR contribution to the BBH merger population from our fiducial model. Points show the fraction $N_{\rm MRR}/N_{\rm BBH}$ in logarithmic metallicity bins, with error bars indicating 95\% Wilson confidence intervals.}
\label{fig:f_MRR}
\end{figure}

Fig.~\ref{fig:f_MRR} presents the fraction of BBH mergers formed through the MRR channel as a function of progenitor metallicity. We find that the MRR fraction increases at lower metallicity, indicating that low-metallicity environments are more favorable for the formation of MRR systems. 
This behavior can be understood as a consequence of weaker stellar winds at low metallicity, which allow the donor star to retain more mass prior to Roche-lobe overflow and facilitate efficient mass transfer onto the secondary star. As a result, the secondary is more likely to accrete sufficient material to exceed the mass of the donor, leading to successful MRR.

\begin{figure*}
    \centering
    \includegraphics[width=0.95\textwidth]{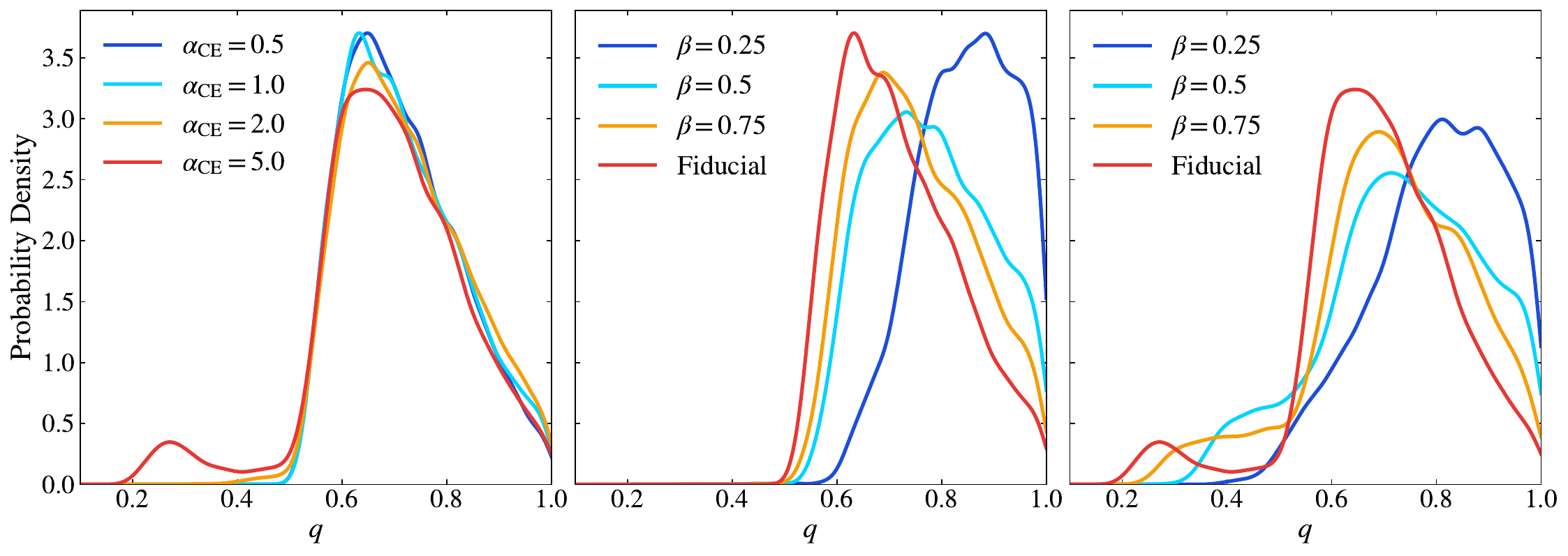}
    \caption{Mass-ratio ($q$) distributions of BBHs formed through the MRR channel for different binary-evolution models. Left: models adopt the fiducial mass-transfer prescription, lognormally distributed natal kicks with $(\mu=5.6)$ and $(\sigma=0.68)$, with different common-envelope efficiencies, $\alpha_{\rm{CE}}$. Middle: models with the same natal-kick prescription,  different mass-transfer prescriptions and assuming $\alpha_{\rm{CE}}=1.0$. Right: the same mass-transfer and natal-kick models as in the middle panel, but for $\alpha_{\rm{CE}}=5.0$.}
    \label{fig:parameter_comp}
\end{figure*}

Fig.~\ref{fig:parameter_comp} shows the mass-ratio distribution of merging BBHs formed through the MRR channel under different binary evolution prescriptions. We first compare models with different $\alpha_{\rm{CE}}$. 
We find that the overall shape of the high-$q$ population is similar across different $\alpha_{\rm{CE}}$ models, with the dominant peak located at $0.6 < q < 0.7$. However, only the model with a relatively large $\alpha_{\rm{CE}}$ ($\alpha_{\rm{CE}}=5.0$) produces an additional population at low mass ratios ($q < 0.5$). 
This low-$q$ component forms a distinct secondary peak that is absent in models with lower $\alpha_{\rm{CE}}$. The origin of this behavior can be understood from the survival of binaries during the CE phase. For systems with extreme mass ratios, unstable mass transfer often leads to a CE phase, and binaries in models with low $\alpha_{\rm{CE}}$ tend to merge before envelope ejection. In contrast, a larger $\alpha_{\rm{CE}}$ allows more efficient envelope ejection and enables these systems to survive as compact BBHs, thereby producing the low-$q$ population.

In our fiducial COMPAS model, the mass-transfer efficiency is determined self-consistently by the thermal timescale of the accretor. To assess the impact of mass-transfer efficiency, we additionally consider models with fixed accretion efficiencies, parameterized by $\beta$, while keeping $\alpha_{\rm{CE}}$ fixed. As shown in the middle panel of Fig.~\ref{fig:parameter_comp}, the fiducial model generally corresponds to relatively conservative mass transfer, yielding an effective accretion efficiency higher than those adopted in the fixed-$\beta$ models. This behavior arises because the thermal-timescale prescription allows the accretor to retain a substantial fraction of the transferred mass whenever it can thermally adjust to the incoming material. For $\alpha_{\rm{CE}}=1.0$ (middle panel), no significant low-$q$ population is produced, irrespective of the adopted mass-transfer prescription. This indicates that variations in mass-transfer efficiency alone are insufficient to form GW241011-like systems when the CE efficiency is low. Nevertheless, increasing $\beta$ systematically shifts the primary peak of the mass-ratio distribution toward smaller values of $q$. This trend is expected, as a higher $\beta$ allows the secondary star to accrete more mass during stable mass transfer, increasing the mass of the second-born BH and thereby reducing the final BBH mass ratio.

For $\alpha_{\rm{CE}}=5.0$ (right panel), the secondary low-$q$ peak reappears and shifts to 
progressively smaller mass ratios with increasing $\beta$. This result demonstrates that the formation of extremely asymmetric BBHs through the MRR channel requires both efficient CE ejection and efficient mass accretion during binary interactions. While the CE efficiency primarily determines whether low-$q$ progenitors survive binary evolution, the mass-transfer efficiency governs the extent of MRR prior to BBH formation.

\section{Conclusions and discussion}\label{con}

GW241011 is characterized by a highly asymmetric mass ratio and a rapidly rotating primary BH. Such systems have often been discussed in the context of hierarchical mergers \citep[e.g.,][]{GW241011,Li2025,He2026}, which provide a natural pathway for producing massive, rapidly rotating BHs. However, recent population analyses suggest that GW241011 is unlikely to have formed in a dense environment such as globular clusters, partly due to the challenges associated with retaining second-generation BHs after gravitational-wave recoil kicks \citep[e.g.,][]{Llobera2026}. Alternatively, the MRR channel provides a natural pathway for achieving both properties. 

Our population synthesis models show that a small fraction ($\sim 0.2\%$) of MRR binaries overlap with the 90\% credible posterior region of GW241011. These systems typically exhibit low mass ratios ($q < 0.5$) and positive effective spins ($\chi_{\rm{eff}} > 0.4$). We find that only models with high CE efficiency ($\alpha_{\rm{CE}} = 5$) can produce such systems, since binaries with lower $\alpha_{\rm{CE}}$ tend to merge during the CE phase before forming compact BBHs. The high-spin primary arises naturally through tidal spin-up in compact binaries that survive the CE stage.

Our analysis indicates that the SMT and CE subchannels play fundamentally different roles in shaping the mass-ratio distribution of MRR systems. The SMT subchannel primarily contributes to the high-q population, with most systems clustering around $q\sim0.7$. 
Increasing the mass-transfer efficiency $\beta$ during the first mass transfer phase allows the accretor to gain more mass, shifting the peak of the mass-ratio distribution towards lower values. However, when $\alpha_{\rm{CE}}$ is low, variations in $\beta$ alone are insufficient to produce systems with $q\lesssim0.5$.

In contrast, the low-$q$ population associated with GW241011-like systems originates almost from the CE-driven MRR channel. These binaries typically experience unstable mass transfer and enter a CE phase, where their survival depends critically on the efficiency of envelope ejection. Our results show that the secondary low-$q$ peak appears only in models with large $\alpha_{\rm CE}$, indicating that efficient CE ejection is required for these extreme systems to avoid merger. The position of this low-$q$ peak is then further modulated by the mass-transfer efficiency, with larger $\beta$ producing progressively smaller mass ratios through a stronger degree of MRR. Therefore, the formation of GW241011-like binaries requires both efficient CE survival and substantial mass accretion during binary interaction, highlighting the coupled roles of $\alpha_{\rm CE}$ and $\beta$ in determining the properties of the MRR population.

The metallicity dependence of the MRR fraction suggests that GW241011-like systems are more likely to originate from low-metallicity environments, where the MRR channel contributes more significantly to the BBH merger population. This result is broadly consistent with recent population studies that identified MRR as an increasingly important formation pathway at low metallicity \citep{Smith2026}. In contrast, dynamical channels are primarily governed by the properties of their host environments rather than binary stellar evolution. Therefore, the different metallicity preferences of isolated and dynamical channels may lead to distinct predictions for the host environments and redshift distribution of GW241011-like systems, which could be tested with future GW observations.

Studies of Wolf-Rayet + O-type star binaries suggest that mass transfer is generally non-conservative, with typical accretion efficiencies falling significantly below the conservative limit ($\beta = 1$) and in some cases as low as $\beta \sim 0.1$ \citep[e.g.,][]{Petrovic2005,Shao2016,Nuijten2025}. However, theoretical studies of early Case A mass transfer in close binaries indicate that tidal locking can allow the secondary to remain below critical rotation, thereby enabling more efficient mass accretion \citep{deMink2007}. In our models, increasing $\beta$ enhances the degree of MRR and shifts the low-$q$ population toward more extreme mass ratios. Consequently, observational constraints on the mass-transfer efficiency provide an important test of the MRR interpretation. If future observations continue to favor strongly non-conservative mass transfer for Wolf-Rayet + O-type star binaries, the formation of GW241011-like systems through the MRR channel may be restricted to a narrower region of parameter space.

The CE-driven MRR model provides a plausible explanation for the high primary spin ($\chi_1$) through tidal synchronization of the He-star progenitor in a tight post-CE orbit. The inferred spin-orbit misalignment of $\theta_1 \sim 31^\circ$ in GW241011, however, poses a challenge to the standard isolated binary evolution scenario and has often been interpreted as evidence favoring a dynamical formation origin \citep{GW241011}. Nevertheless, an isolated binary origin cannot be ruled out if BHs receive appreciable natal kicks during their formation, which can tilt the orbital plane relative to the BH spin axes \citep[e.g.,][]{Kalogera2000}. Such kicks may arise from asymmetric mass ejection \citep[e.g.,][]{Janka2013} or neutrino-driven mechanisms during core collapse \citep[e.g.,][]{Fryer2006}. Alternatively, the spin axes of newly formed BHs may be intrinsically reoriented during core collapse through the spin-axis tossing mechanism proposed by \citet{Tauris2022}. Furthermore, the MRR channel can reproduce several of the key observed properties of GW241011, including its extreme mass ratio and rapidly spinning primary BH. Therefore, although the inferred spin-orbit misalignment remains an important challenge for isolated binary evolution, it is not, by itself, sufficient to rule out an isolated binary origin.

Unlike dense stellar environments such as globular clusters, young massive clusters, and nuclear star clusters, active galactic nucleus (AGN) disks have emerged as a promising environment for producing low-mass-ratio BBH mergers. For instance, \citet{Vaccaro2026} found that the dynamics of AGN disks can naturally produce low-mass-ratio mergers, as repeated hierarchical mergers and gas-driven pairing mechanisms can facilitate mergers between massive merger remnants and lower-mass companions. However, it remains unclear whether the AGN-disk channel can reproduce the observed properties of GW241011.

\begin{acknowledgments}
We thank the anonymous referee for helpful suggestions and Yong Shao and Yihan Wang for useful discussions and comments. Y.Q. acknowledges support from the National Natural Science Foundation of China (grant Nos. 12473036 and 12573045) and partial support from the Jiangxi Provincial Natural Science Foundation (grant No. 20242BAB26012). This work was supported by computational resources provided by Expanse \citep{strande2021expanse}.
\end{acknowledgments}

\software{
          \texttt{COMPAS} {\citep[version 03.27.02;][]{Stevenson2017,Vigna2018,Neijssel2019,TeamCOMPAS2022,TeamCOMPAS2025}};  \texttt{Python}, \url{https://www.python.org}, \texttt{NumPy} \citep{harris2020}, \texttt{Matplotlib} \citep{Hunter2007}, \texttt{seaborn} \citep{Waskom2021}.          
          }

\bibliography{main}{}
\bibliographystyle{aasjournalv7}

\end{document}